\documentclass[pdflatex,sn-mathphys-num]{sn-jnl}
\newcommand{\lp}{\left(}
\newcommand{\rp}{\right)}
\newcommand{\lb}{\left[}
\newcommand{\rb}{\right]}

\newcommand{\non}{\nonumber}

\newcommand{\intd}{\int\displaylimits}	
\newcommand{\sumd}{\sum\displaylimits}
\newcommand{\be}{\begin{equation}} 
\newcommand{\ee}{\end{equation}} 
\newcommand{\bea}{\begin{eqnarray}} 
\newcommand{\eea}{\end{eqnarray}} 
\newcommand{\bm}{\begin{mathletters}} 
\newcommand{\eml}{\end{mathletters}}

\usepackage{graphicx}%
\usepackage{multirow}%
\usepackage{amsmath,amssymb,amsfonts}%
\usepackage{amsthm}%
\usepackage{mathrsfs}%
\usepackage[title]{appendix}%
\usepackage{xcolor}%
\usepackage{textcomp}%
\usepackage{manyfoot}%
\usepackage{booktabs}%
\usepackage{algorithm}%
\usepackage{algorithmicx}%
\usepackage{algpseudocode}%
\usepackage{listings}%
\usepackage{physics} 

\theoremstyle{thmstyleone}%
\theoremstyle{thmstyletwo}%

\theoremstyle{thmstylethree}%

\begin{document}

\title{Density--Dependent Description of Clustering in 
       Axially--Symmetric Nuclei}


\author*[1]{\fnm{A.} \sur{Dumitrescu}}
\email{alexandru.dumitrescu@theory.nipne.ro}

\author[1,2]{\fnm{D.S.} \sur{Delion}}\email{delion@theory.nipne.ro}

\affil*[1]{\orgdiv{Department of Theoretical Physics}, 
\orgname{Horia Hulubei National Institute for R\&D in Physics
and Nuclear Engineering}, \orgaddress{\street{Reactorului 30}, 
\city{Magurele}, \postcode{077125}, \state{Ilfov}, \country{Romania}}}

\affil[2]{\orgdiv{Department of Physical Sciences}, 
\orgname{Academy of Romanian Scientists}, \orgaddress{\street{Ilfov 3}, 
\city{Bucharest}, \postcode{050044}, \state{sector 5}, 
\country{Romania}}}


\abstract{We present a microscopic description of the $\alpha$-decay 
processes from axially-symmetric superfluid nuclei in terms of proton 
($\pi$) and neutron ($\nu$) single--particle (sp) degrees of freedom 
within a standard Woods--Saxon mean field (WSMF) and a 
density--dependent clustering potential centered around the average Mott 
point of the nuclear density, where the $\alpha$--particle is formed. In 
this way the clustering process is automatically enhanced on the nuclear 
surface. As an illustration of the method, a systematics of 
$\alpha$--decay from even-even emitters is presented.}


\maketitle

\section{Introduction}\label{sec1}

The phenomenon of radioactivity has been known for well over a century 
\cite{Bec86,Bec86a}. The theory of $\alpha$--decay was established by 
Gamow \cite{Gam28} and independently by Gurney \& Condon \cite{Gur28}, 
and it represented the first application of the probabilistic 
interpretation of quantum mechanics. This field of research has produced
many interesting theoretical and experimental results. Of the latter,
fairly recently it was empirically proven that nucleonic clusters are
present within the structure of finite nuclear systems even when decay
is energetically forbidden \cite{Tan21}. Historical notes on the topic 
of radioactivity and a comprehensive overview of the theoretical 
research and some experimental aspects investigated over the decades can 
be found in Refs. \cite{Lov98,Del15,Del18,Dum22,Dum25}. This work is 
mainly concerned with the still--developing microscopic theory of 
emission processes, focusing particularly on the clustering phenomenon 
in terms of nucleonic degrees of freedom. This line of investigation
concerning finite nuclei began in the 1960s \cite{San62,Man64,Pog69} 
based on earlier results of Lane \cite{Lan60}. In the 
1990s the problem of nucleonic clustering was studied extensively within
infinite nuclear matter \cite{Rop98}, while modern approaches to finite 
systems along with many interesting developments took place since the 
early 2000s for light \cite{Fre18}, medium mass or heavy nuclei 
\cite{Das23,Bet12,Rop14}. Interesting new methods have been developed,  
as for instance nonlinear quantum dynamics on droplets \cite{Car21}. 

Here we continue investigations started at the spherical mean--field 
(MF) level by Delion and Liotta \cite{Del13} and developed further in a 
self--consistent approach by Dumitrescu and Delion for spherical nuclei 
\cite{Dum23,Dum25}. Self--consistent applications to axially--symmetric 
nuclei will be presented in the future. In the present work we forego 
the requirement of self--consistency and develop a simpler theory that 
produces similar results. The discussion is organized as follows. 
Section \ref{sec2} covers the formal developments of our 
density--dependent theory for axially--symmetric clustered nucleonic 
systems, together with a calculation of the cluster formation amplitude. 
Section \ref{sec3} is concerned with applications to open shell nuclei 
across the range of known even--even alpha-emitters. Section \ref{sec4} 
presents our conclusions.

\section{Theory}\label{sec2}

We begin by developing the sp potential where the $\alpha$--particle is 
formed and then we discuss the calculation of the decay width. 

\subsection{Cluster Density--Dependent Interaction} 

The underlying physics is that of a standard axially--symmetric WSMF 
containing spin--orbit and Coulomb terms. We work in the universal 
parametrization \cite{Dud82,Cwi87} and the field is expanded in monopole
and quadrupole terms. Residual pairing interactions are treated within 
the usual Bardeen--Cooper--Schrieffer (BCS) approximation. To this
framework we add a Cluster Density--Dependent Interaction (CDDI) that
describes the formation of $\alpha$--particles at nuclear densities
$\rho\approx\rho_{\textrm{Mott}}$, where we consider the prescription
for the Mott point corresponding to 10\% of the equilibrium density
of symmetric nuclear matter \cite{Rop98}. This interaction is defined as
\bea
V\lp{\bf r},{\bf r}^{\prime}\rp&=&-v_{0}e^{-\frac{\lvert{\bf r}-
{\bf r}^{\prime}\rvert^{2}}{b^{2}}}\lb 1+x_{\textrm{c}}
V_{\textrm{clus}}\lp{\bf R}\rp\rb\non\\
V_{\textrm{clus}\lp{\bf R}\rp}&=&
e^{-\frac{\lb\rho\lp{\bf R}\rp-\rho_{\textrm{Mott}}\rb^{2}}{B^{2}}}.
\eea
The first term describes the relative interaction between a pair
of particles, the effective length of the interaction being $b$. We
have shown in our previous works \cite{Dum23,Dum25} that when integrated
over the nuclear density, this term reproduces the central part of the 
WSMF. The second term includes a phenomenological interaction between a 
pocket of nuclear matter located at the center of mass (com) of the pair 
at radius ${\bf R}$ and the surface regions of the nucleus around the 
Mott point, where we assume implicitly that the interaction is along the 
radius vector of the com. Here, $B$ denotes the clustering width in 
terms of a fraction of the equilibrium density. The interaction 
constants $v_{0}$ and $x_{\textrm{c}}$ only play a part in the theory as 
the product 
\bea
v_{\textrm{clus}}=v_{0}x_{\textrm{c}},
\eea
given that the only modification to the standard WSMF comes from the 
surface term $V_{\textrm{clus}}\lp{\bf R}\rp$. This follows from 
approximating the relative part of the interaction by 
$\delta\lp{\bf r}-{\bf r}^{\prime}\rp$. The approximation holds because
typically such a Gaussian, when folded over nucleonic density, will 
accurately reproduce a WSMF when $b$ is small. Furthermore, in this 
context the radius of the com close to the nuclear surface can also be 
approximated 
as 
\bea
{\bf R}=\frac{{\bf r}+{\bf r}^{\prime}}{2}\approx{\bf r}
\eea
due to the sharpness of the surface. In previous works where a 
self--consistent treatment was pursued \cite{Dum25}, both terms and the 
angular integrals involved in the determination of ${\bf R}$ were 
evaluated exactly. Qualitatively, both this exact treatment and the 
above approximation result in the creation of a Gaussian--shaped local 
minimum of the potential in the region of the nuclear surface which 
serves to enhance nucleonic clustering in that area. 
We considered here the mean nuclear density
\bea
\rho\lp{\bf r}\rp=\frac{\rho_{\pi}\lp{\bf r}\rp+
\rho_{\nu}\lp{\bf r}\rp}{2}
\eea
where each component is given by the standard relation and multipole 
expansion
\bea
\rho_{\tau}\lp{\bf r}\rp=\sumd_{m}v_{\tau m}^{2}\lvert\Phi_{\tau m}
\lp\bf r\rp\rvert^{2}=\sumd_{L}\frac{\sqrt{4\pi}}{\hat{L}}
\rho_{\tau L}\lp r\rp Y_{L0}\lp\theta\rp,~\tau=\pi,\nu.
\eea
Here $v_{\tau m}$ are the BCS occupation amplitudes and $\Phi_{\tau m}$ 
are the Nilsson eigenfunctions of total angular momentum intrinsic 
projection  $m$ for protons ($\tau=\pi$) and neutrons ($\tau=\nu$). 
The internal spin degrees of freedom are orthogonalized 
trivially and hence the density depends only on spatial coordinates.
$\rho_{\textrm{Mott}}$ is generally taken as
\bea
\rho_{\textrm{Mott}}=a\rho\lp 0\rp,~a\in\lp 0.1,0.2\rp.
\eea
This interaction is automatically maximal on the nuclear surface
and has a larger tail beyond it. Furthermore, the cluster term is local
in this form, as the exchange contribution vanishes. Then, with 
\bea
V_{\textrm{clus}}\lp{\bf r},{\bf r}^{\prime}\rp\approx
\delta\lp{\bf r}-{\bf r}^{\prime}\rp 
e^{-\frac{\lb\rho\lp{\bf R}\rp-\rho_{\textrm{Mott}}\rb^{2}}{B^{2}}}
\eea
the direct term of the cluster Hartree--Fock field is
\bea
U_{\textrm{clus}}\lp{\bf r}\rp=\intd_{\mathcal{D}^{\prime}}
V_{\textrm{clus}}\lp{\bf r},{\bf r}^{\prime}\rp
\rho\lp{\bf r}^{\prime}\rp\dd{\bf r}^{\prime}=v_{\textrm{clus}}
e^{-\frac{\lb\rho\lp{\bf r}\rp-\rho_{\textrm{Mott}}\rb^{2}}{B^{2}}}
\rho\lp{\bf r}\rp
\eea
and expands in multipoles similarly to the density.
Thus, the dependence on deformation is automatically taken into account
by the mean nuclear density.

\subsection{Decay Width}

A thorough overview of all computational details pertaining to this
section can be found in a monography by Delion \cite{Del10}. The main
result is a sum of partial widths according to the $\alpha$--particle's
angular momentum
\bea
\Gamma=\sumd_{L_{\alpha}}\Gamma_{L_{\alpha}}=\sumd_{L_{\alpha}}\hbar
v_{\alpha}\lvert N_{L_{\alpha}}\rvert^{2}
\eea
where $v_{\alpha}=\sqrt{\frac{2Q_{\alpha}}{\mu_{\alpha}}}$ is the 
asymptotic $\alpha$--particle velocity depending on the Q-value of the 
process $Q_{\alpha}$ and the reduced mass of the $\alpha$--daughter 
system $\mu_{\alpha}$. The quantities $N_{L_{\alpha}}$ are channel
scattering amplitudes found by matching the internal and external 
components of the $\alpha$--particle wave function
\bea\label{match}
f_{L_{\alpha}}^{\lp\textrm{int}\rp}\lp R_{\alpha}\rp=
f_{L_{\alpha}}^{\lp\textrm{ext}\rp}\lp R_{\alpha}\rp
\eea
which are calculated in the $\alpha$--daughter interaction potential.
The external component is a superposition of fundamental solutions with
the outgoing Coulomb--Hankel asymptotics 
\bea
f_{L_{\alpha}}^{\lp\textrm{ext}\rp}\lp R_{\alpha}\rp=
\sumd_{L_{\alpha}^{\prime}}\mathcal{H}_{L_{\alpha} 
L^{\prime}_{\alpha}}^{\lp+\rp}
\lp R_{\alpha}\rp N_{L_{\alpha}^{\prime}}
\xrightarrow[R_{\alpha}\rightarrow\infty]{}H_{L_{\alpha}}^{\lp +\rp}
\lp\chi,\kappa R_{\alpha}\rp
\eea
involving the Coulomb parameter $\chi=\frac{4Z_{\textrm{D}}}
{\hbar v_{\alpha}}$ and reduced radius $\kappa R_{\alpha}$ with
$\kappa=\mu_{\alpha}v_{\alpha}$. $\mathcal{H}_{L_{\alpha} 
L^{\prime}_{\alpha}}^{\lp+\rp}$ is found through backward integration
and the channel scattering amplitudes follow from the inversion
of the matrix according to Eq. (\ref{match})
\bea
N_{L_{\alpha}}=\sumd_{L_{\alpha}^{\prime}}\lb\mathcal{H}_{L_{\alpha} 
L^{\prime}_{\alpha}}^{\lp+\rp}\rb^{-1}
f_{L_{\alpha}^{\prime}}^{\lp\textrm{int}\rp}\lp R_{\alpha}\rp.
\eea
It should be noted that in the semiclassical approximation, for a pure 
Coulomb field one can construct the matrix of fundamental solutions
according to the Fr\"{o}man method \cite{Fro57}. The internal function
is called $\alpha$--particle formation amplitude and is given by the 
overlap between parent and product of daughter and $\alpha$--particle 
wave functions 
\bea
f^{\lp\textrm{int}\rp}\lp{\bf R}_{\alpha}\rp=
\langle\psi_{\textrm{P}}\vert\psi_{\textrm{D}}\psi_{\alpha}\rangle.
\eea
Around the geometrical contact radius 
$\overline{R}_{\textrm{C}}=1.2A_{\textrm{D}}^{\frac{1}{3}}$, 
antisymmetrization effects are small enough that they can be neglected 
in the $\alpha$--daughter product. The multipole expansion of such a 
function is
\bea
f^{\lp\textrm{int}\rp}_{L_{\alpha}}\lp R_{\alpha}\rp=\sumd_{N_{\alpha}}
W_{N_{\alpha}L_{\alpha}}
\mathcal{R}_{N_{\alpha}L_{\alpha}}^{\lp 4\beta\rp}\lp R_{\alpha}\rp
\eea
with the W coefficients given in Chapter 9.6.1 of \cite{Del10} in
terms of Moshinsky symbols, $uv$ products of BCS amplitudes and sp 
Nilsson expansion coefficients in the spherical harmonic oscillator 
basis.

\section{Application}\label{sec3}

\begin{figure}
\centering
\includegraphics[width=\columnwidth]{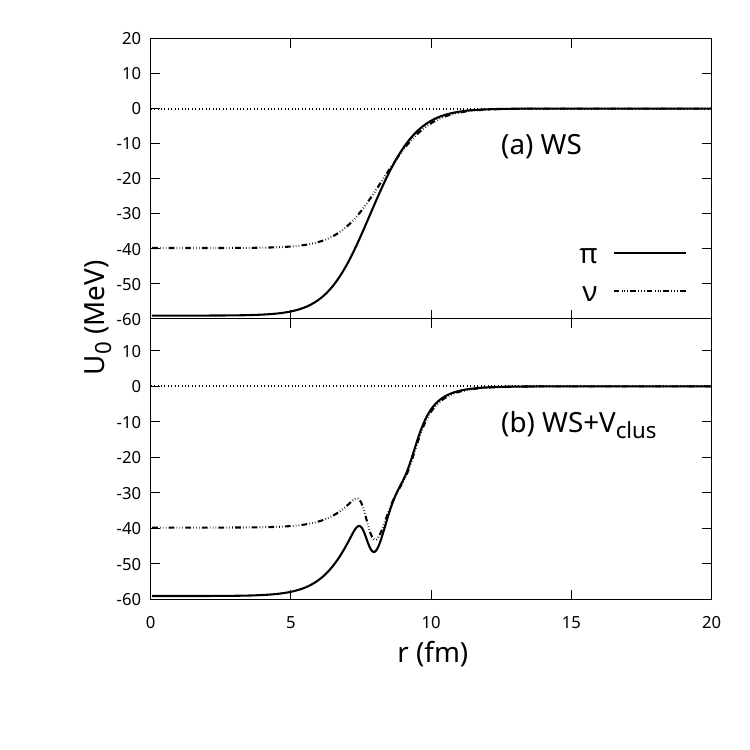} 
\caption{Monopole fields for the standards WS case (panel a) and
WS plus cluster potential (panel b) for protons and neutrons in the
case of $^{242}\textrm{Pu}$.}
\label{fig1}
\end{figure}

The theory presented in the previous section was used to study 
$\alpha$--decay data from the ground state of even--even emitters. All
decay data were taken from the AME2020 table \cite{Wan21}. The 
quadrupole deformations were taken from the FRDM table \cite{Mol16}. 
Pairing gaps were calculated according to the 5-point formula of Ref. 
\cite{Ben00}. The $\alpha$--daughter interaction potential was 
calculated within the double folding procedure of the M3Y interaction 
\cite{Ber77,Car92}. We first show some illustrative results for the 
$\alpha$--decay of $^{242}\textrm{Pu}$. Figure \ref{fig1} shows the 
monopole component of the MF potential in the standard WS case in panel 
(a) and WS plus cluster potential in panel (b), separately for protons 
(continous line) and neutrons (dashed line) as a function of radius. The 
interaction constant reproducing the experimental decay width is 
$v_{\textrm{clus}}=313~\textrm{MeV}$, while both $\rho_{\textrm{Mott}}$ 
and $B$ were taken as $10\%$ of the central density. In order for the 
resulting surface minima to overlap spatially and thus enhance nucleonic 
clustering, an averaging procedure is used over the proton and neutron 
densities so that an averaged Mott radius results. Otherwise, the effect
on clustering is very weak.   

\begin{figure}
\centering
\includegraphics[width=\columnwidth]{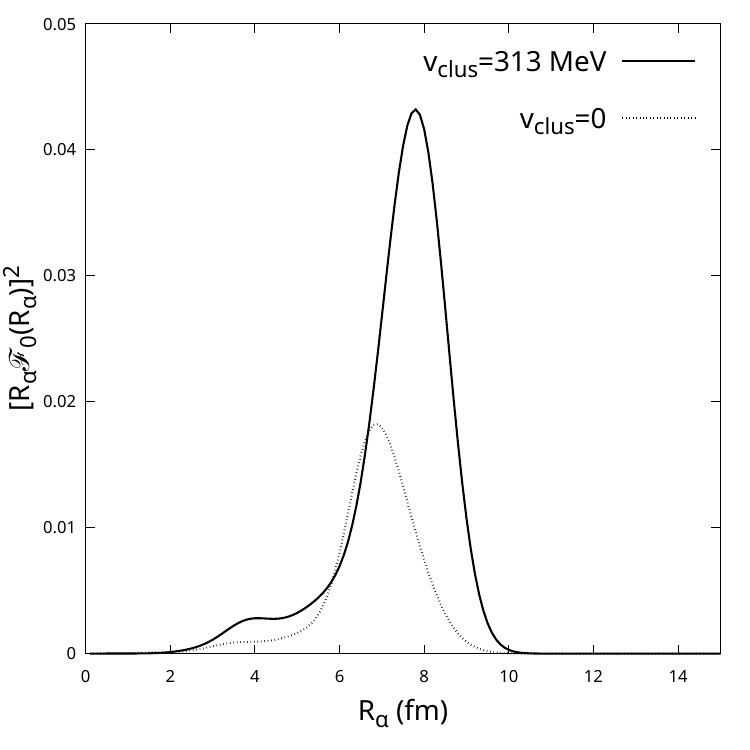} 
\caption{Monopole of the $\alpha$--particle formation amplitude as a 
function of the particle's com for the standard WS case (dashed line) 
and WS plus cluster potential case (continuous line) for 
$^{242}\textrm{Pu}$.}
\label{fig2}
\end{figure}

Figure \ref{fig2} shows the monopole part of the formation amplitude as 
a function of the $\alpha$--particle com in the WS (dashed line) and WS 
plus cluster potential (continuous line) cases. One observes an 
enhancement of about a factor 2 in the peak of the formation amplitude 
as well as a shift of the peak towards the location of the two minima in 
the sp potentials.

\begin{figure}
\centering
\includegraphics[width=\columnwidth]{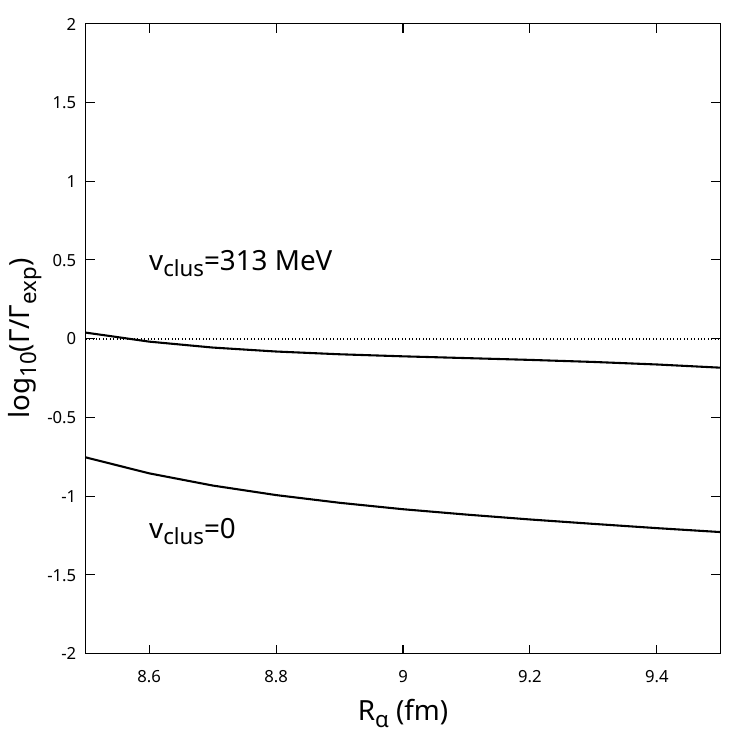} 
\caption{Exact calculation of the decimal logarithm of theoretical to 
experimental decay widths ratio as a function of the $\alpha$--particle 
com radius.}
\label{fig3}
\end{figure}

Figure \ref{fig3} illustrates the decimal logarithm of the ratio between
calculated and observed decay width for $^{242}\textrm{Pu}$ versus
$\alpha$--particle com radius using the exact propagator for the WS and 
WS plus cluster potential case. Notice that the exact procedure result 
is quite stable and that the cluster strength is required in order to 
remove roughly an order of magnitude between the theoretical and 
experimental results.

\begin{figure}
\centering
\includegraphics[width=\columnwidth]{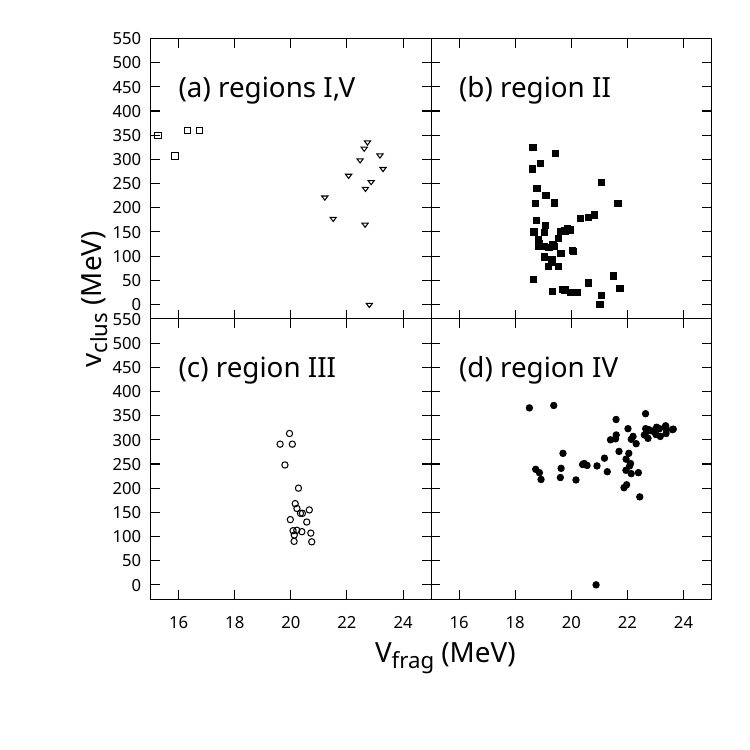} 
\caption{Clustering strength versus fragmentation potential for the
five magic regions.}
\label{fig4}
\end{figure}

\begin{figure}
\centering
\includegraphics[width=\columnwidth]{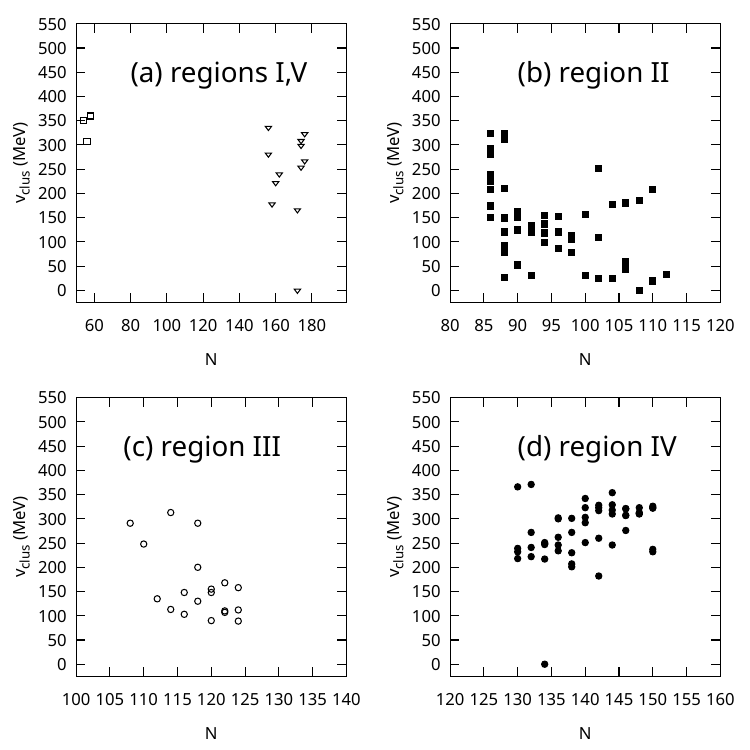} 
\caption{Clustering strength versus neutron number for the
five magic regions.}
\label{fig5}
\end{figure}

Figure \ref{fig4} presents the systematics of the strength 
$v_{\textrm{clus}}$ which reproduces the observed decay width in each 
case as a function of the fragmentation potential $V_{\textrm{frag}}$ 
defined as the difference between the height of the $\alpha$--daughter 
potential at the barrier radius and $Q_{\alpha}$, where the panels 
depict magic regions according to Table \ref{tab1}. We observe for each 
region a linear correlation between the two quantities, showing that the 
clustering strength is related to the $\alpha$--particle reduced decay 
width introduced in the factorization
\bea
\Gamma_{0}=2\gamma_{0}^{2}P_{0}\lp R_{\textrm{B}}\rp
\eea
where $P_{0}$ is the monopole penetrability and $R_{\textrm{B}}$
is the barrier radius. Thus, the clustering strength behaves similarly
to the universal decay rule for the reduced width \cite{Del09}. 
Similarly, Figure \ref{fig5} shows the clustering strength versus 
neutron number for the five magic regions, with clear linear
correlations that peak in the vicinity of magic neutron numbers.

\begin{table}[h]
\caption{Magic regions of the nuclear chart}\label{tab1}
\begin{tabular}{@{}lll@{}}
\toprule
Region & Z & N \\
\midrule
I    & $<82$   & $\in\lb 50,82\rp$   \\
II    & $<82$   & $\in\lb 82,126\rp$  \\
III    & $\ge 82$ & $\in\lb 82,126\rp$  \\
IV  & $\ge 82$ & $\in\lb 126,152\rp$ \\ 
V & $\ge 82$ & $\ge 152$ \\
\botrule
\end{tabular}
\end{table}

\section{Conclusion}\label{sec4}

We presented a microscopic theory of cluster emission processes for 
axially-symmetric superfluid nuclei involving a standard Woods--Saxon 
mean field (WSMF) and a density--dependent clustering potential centered 
around the average Mott point of the nuclear density, thus enhancing the 
clustering process on the nuclear surface. The multipoles of the 
$\alpha$--particle formation amplitude were calculated using the exact 
propagator, this approach generally yielding a good result. Experimental 
decay widths were reproduced using the clustering strength 
$v_{\textrm{clus}}$ and it was shown that this parameter is linearly 
related both to the fragmentation potential $V_{\textrm{frag}}$ and the 
neutron number $N$ across the nuclear chart in the case of even--even 
emitters.

\backmatter

\bmhead{Acknowledgements}

This work was supported by a grant of the Ministry of Research, 
Innovation and Digitization, CNCS--UEFISCDI, Project No. 
PN--IV--P1--PCE--2023--0273, within PNCDI IV, and from the grant having
Project No. PN--23--21--01--01/2023.

\section*{Declarations}

\begin{itemize}
\item Funding: see Acknowledgements.
\item Conflict of interest/Competing interests: Not applicable
\item Ethics approval and consent to participate: Not applicable
\item Consent for publication: Not applicable
\item Data availability : all data are available in the sources 
mentioned in the bibliography.
\item Materials availability: not applicable.
\item Code availability: for coding details, the authors can provide 
information upon reasonable request. 
\item Author contribution: the idea of the CDDI approach and most of the
code in use were developed by D.S. Delion. Some refinements related to
the averaging of Mott point radii, debugging of the software, the 
calculations and manuscript drafting were caried out by A. Dumitrescu.
\end{itemize}


\bibliography{nuclear}

\end{document}